# All-electrical switching of a topological non-collinear antiferromagnet at room temperature

Yongcheng Deng,[1,2#] Xionghua Liu,[1,2#] Yiyuan Chen,[3,4#] Zongzheng Du,[3,4#] Nai Jiang,[1,2] Chao Shen,[1,2] Enze Zhang,[1,2] Houzhi Zheng,[1,2] Hai-Zhou Lu,[3,4*] Kaiyou Wang[1,2,5,6*]

[1] *State Key Laboratory for Superlattices and Microstructures, Institute of Semiconductors, Chinese Academy of Sciences, Beijing 100083, China*

[2] *Center of Materials Science and Optoelectronics Engineering, University of Chinese Academy of Sciences, Beijing 100049, China*

[3] *Institute for Quantum Science and Engineering and Department of Physics, Southern University of Science and Technology (SUSTech), Shenzhen 518055, China*

[4] *Shenzhen Key Laboratory of Quantum Science and Engineering, Shenzhen 518055, China*

[5] *Beijing Academy of Quantum Information Sciences, Beijing 100193, China*

[6] *Center for Excellence in Topological Quantum Computation, University of Chinese Academy of Science, Beijing 100049, China*

[#] These authors contributed equally to this work.

[*] Correspondence and requests for materials should be addressed to H.L., or K.W.

[*] Email: luhz@sustech.edu.cn, kywang@semi.ac.cn

**Non-collinear antiferromagnetic Weyl semimetals, combining the advantages of a zero stray field and ultrafast spin dynamics as well as a large anomalous Hall effect and the chiral anomaly of Weyl fermions, have attracted extensive interests. However, the all-electrical control of such systems at room temperature, a crucial step toward practical applications, has not been reported. Here using a small writing current of around $5 \times 10^6$ A cm$^{-2}$, we realize the all-electrical current-induced deterministic switching of the non-collinear antiferromagnet Mn$_3$Sn with a strong readout signal at room temperature in the Si/SiO$_2$/Mn$_3$Sn/AlO$_x$ structure, without external magnetic field and injected spin current. Our simulations reveal that the switching is originated from the current-induced intrinsic non-collinear spin-orbit torques in Mn$_3$Sn itself. Our findings pave the way for the development of topological antiferromagnetic spintronics.**

Antiferromagnets have recently attracted tremendous interests as candidates for next-generation spintronics devices with the prospect of offering higher storage density and faster data processing than their ferromagnetic counterparts[1,2]. However, the weak readout signals of conventional collinear antiferromagnets driven by electrical approaches greatly restrict their practical applications[3,4]. Alternatively, large magnetotransport signatures such as the intrinsic anomalous Hall effect in topological antiferromagnets could provide a solution to this issue[5,6]. In particular, the non-collinear antiferromagnetic Weyl semimetal $Mn_3Sn$ has recently fascinated the condensed matter physics and information technology communities because of its nontrivial band topology[7,8] and unusual magnetic responses[9-11].

$Mn_3Sn$ hosts an ABAB stacking sequence of the (0001) kagome lattice of Mn (Fig. 1a) with a 120° non-collinear antiferromagnetic ordering of the Mn magnetic moments below the Néel temperature of $T_N \approx 430$ K[9]. This antiferromagnetic state on the kagome bilayers can be viewed as a ferroic ordering of a cluster magnetic octupole (Fig. 1b), which macroscopically breaks the time-reversal symmetry and results in a large anomalous Hall effect[9,12]. With the assistance of an auxiliary magnetic field, the deterministic switching of the magnetic octupole in $Mn_3Sn$ has been achieved by spin-orbit torques from a heavy-metal layer[13,14]. However, as a critical step toward practical applications, the field-free manipulation of $Mn_3Sn$ driven by electrical currents at room temperature has not been reported. Here, we demonstrate the all-electrical switching of the topological antiferromagnetic states in heavy-metal-layer-free $Mn_3Sn$ devices.

Experiments were performed on sputter-deposited $Mn_3Sn$ (50 nm)/$AlO_x$ (2 nm) thin films, and the reference samples with heavy metals consisting of Ru (3 nm)/$Mn_3Sn$ (50 nm)/$AlO_x$ (2 nm) and Ru (3 nm)/$Mn_3Sn$ (50 nm)/Pt (8 nm)/$AlO_x$ (2 nm). All samples were deposited on thermally oxidized Si substrate. Unless otherwise stated, all the measurements were performed on the $Mn_3Sn$ (50 nm) /$AlO_x$ (2 nm) materials or devices without heavy metals. We first characterize the structure, transport properties, and magnetic properties of $Mn_3Sn$. The X-ray diffraction peaks of (010) and (020) at 18 and 36 degrees confirm the hexagonal $D0_{19}$ $Mn_3Sn$ structure, and no additional peaks coming from plausible impurity phases were observed[15,16] (see the details in Supplementary Section S1 and more related analysis in Supplementary Sections S2-S4). Besides, the microstructure of our film and the chemical composition of $Mn_{3.06}Sn_{0.94}$ were measured by cross-sectional high-resolution transmission electron microscopy (HR-TEM) and energy-dispersive X-ray spectroscopy (EDX)[17,18] (Supplementary

Fig. S3), respectively. The magnetotransport phenomena observed in our thin films (Fig. 1c and Supplementary Fig. S4) are consistent with the previous measurements[13], where the angular dependence of the in-plane longitudinal magnetoconductivity ($\Delta\sigma = \sigma - \sigma_\perp$) and planar Hall conductivity $\sigma_H^{PHE}$ can be well fitted by the theoretical equations for the chiral anomaly of Weyl fermions[19] (Fig. 1c and Supplementary Section S6).

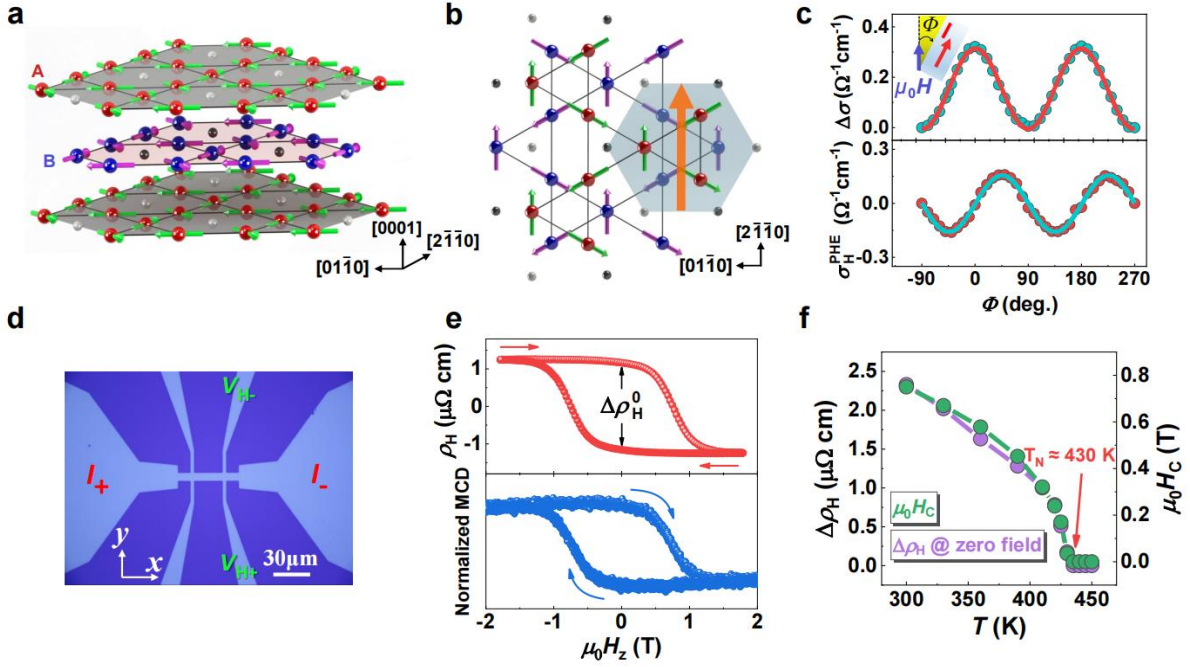

**Fig. 1 | Crystal and magnetic structures, magnetotransport and magnetic properties of the Mn$_3$Sn device**. **a**, Mn$_3$Sn crystal structure. The large red and blue spheres (small grey and black spheres) represent the Mn (Sn) atoms at the $z = 0$ and 1/2 planes, respectively. **b**, Magnetic structure of Mn$_3$Sn. The Mn magnetic moments (green and pink arrows) lie in the kagome-plane and form an inverse triangular spin structure, ferroic ordering of a cluster magnetic octupole (big orange arrow) made of six moments (colored hexagon). **c**, Angular dependence of the in-plane longitudinal magnetoconductivity $\Delta\sigma$ (top) and planar Hall conductivity $\sigma_H^{PHE}$ (bottom) of the 50-nm-thick Mn$_3$Sn device at room temperature and 1.8 Tesla. The red and cyan solid lines for $\Delta\sigma$ and $\sigma_H^{PHE}$ are the fitting results using the theoretical equations for the chiral anomaly (Eqs. S6-S7 in Supplementary Section S6). **d**, Optical micrograph of our fabricated Hall device and measurement scheme. **e**, (top) Anomalous Hall resistivity and (bottom) magnetic circular dichroism (MCD) signal versus $\mu_0 H_z$ for the 50-nm-thick Mn$_3$Sn device at room temperature. Both of them exhibit clear hysteresis loops with a

coercive field $\mu_0 H_C$ of approximately 0.75 T. **f**, Temperature dependence of $\Delta\rho_H^0$ and $\mu_0 H_C$ derived from the hysteresis loops of the 50-nm-thick Mn$_3$Sn device, suggesting a Néel temperature $T_N \approx 430$ K (Supplementary Fig. S7).

We measured the anomalous Hall resistivity $\rho_H$ as a function of the out-of-plane magnetic field $\mu_0 H_Z$ to quantitatively estimate the population of switchable domains in the device (Fig. 1d). A clear hysteresis of the anomalous Hall resistivity with a zero-field change $\Delta\rho_H^0$ is observed (Fig. 1e (top)), which shows that the negative (positive) value of $\rho_H$ is produced by the '+z ($-z$) domain with the positive $+z$ (negative $-z$) component of the polarization direction of the magnetic octupole[13,15,20,21]. The field-swept measurements of the magnetic circular dichroism (MCD) exhibit a hysteresis loop of $M$ - $\mu_0 H_Z$ with a coercive field of $\mu_0 H_C \approx 0.75$ T (Fig. 1e (bottom)), which is in agreement with the measured anomalous Hall resistivity (Fig. 1e (top)). Moreover, the $M$ - $\mu_0 H_Z$ curves were measured at different temperatures ($T$) using a vibrating sample magnetometer and the extracted magnetization of around 8 emu/cc at 300 K and 5 mT (Supplementary Fig. S5f), which is comparable with the previous reported values[13,15,20,21]. The longitudinal resistivity and Hall resistivity as a function of temperature under zero magnetic field are illustrated in Supplementary Figs. S6. Both the $M$ - $T$ and $\rho_H$ - $T$ curves show a rapid decrease at around 250 K, which corresponds to the transition to spiral phase[22-24]. Besides, the Néel temperature $T_N$, corresponding to the disappearance of $\Delta\rho_H^0$ and $\mu_0 H_C$ of the anomalous Hall hysteresis loops, is found to be approximately 430 K (Fig. 1f and Supplementary Fig. S7), which is close to that of single crystal Mn$_3$Sn[9]. The results confirm that our thin films have similar physical properties to those of previous reports.

We then examine the possible current-induced topological non-collinear antiferromagnetic state switching. For the reference samples, the current-induced deterministic switching can only be observed under an auxiliary magnetic field for Ru/Mn$_3$Sn/Pt devices (Supplementary Fig. S8), which are consistent with the previous works[13,14]. Interestingly, different current-induced switching behaviors were observed in our heavy-metal-layer-free devices. Figure 2a presents the $\rho_H - \mu_0 H_Z$ curve of a 50-nm-thick Mn$_3$Sn device. Under zero magnetic field, a $50-$ ms writing current pulse $I_{write}$ followed by a DC reading current of $I_{read} = 0.1$ mA is applied along the $x$ direction. Surprisingly, as shown the black curve in Fig.

2b, the electrical current flowing through the device leads to a clear negative (positive) jump in $\rho_H$ at a positive (negative) threshold writing current, implying a reversing of the $z$ component of the octupole. The magnitude of the Hall resistivity jump $\Delta\rho_H^J$ is approximately 58% of $\Delta\rho_H^0$ in the field-swept measurements (Fig. 2a). Figure 2b also shows the $\rho_H - J_{write}$ loops in an in-plane magnetic field $\mu_0 H_x$ of $\pm$ 0.2 T and $\pm$ 0.4 T. With the increase of positive (negative) applied magnetic field, the gradual shift of $\rho_H - J_{write}$ loops toward a negative (positive) $J_{write}$, together with the reduction of $\Delta\rho_H^J$, is observed, which is clearer in the field dependence of the threshold current in Fig. 2c. The current-induced magnetization switching disappears for a sufficiently large bias field, e.g., 1 T > $\mu_0 H_C$, probably because the magnetic octupoles are aligned to the external field direction.

Compared with the Mn$_3$Sn/(heavy-) metal reference devices (Supplementary Fig. S8), here, the writing current is fully injected into the Mn$_3$Sn layer without passing through a highly conductive metal layer. In addition, the strong inversion asymmetry along the $x(y)$ direction was confirmed by our nonlinear Hall measurements[25,26]. As a low frequency AC current is applied along the $x(y)$ direction, a considerable second harmonic voltage is measured along the $y(x)$ direction in our Mn$_3$Sn device, probably due to the boundaries between grains and/or magnetic domains. The nonlinear Hall effect in our Mn$_3$Sn device is considerably stronger than that measured in the reference samples, justifying a larger Rashba effect in the heavy-metal-layer-free device (Supplementary Fig. S9). We explain that the field-free deterministic switching originates from current-induced spin accumulations in the Mn$_3$Sn layer without inversion symmetry.

Notably, the critical writing current density $J_c$ required for switching the octupole, at zero magnetic field, is estimated to be $5 \times 10^6$ A cm$^{-2}$ in our Mn$_3$Sn device, which is less than half required for the reference Ru/Mn$_3$Sn/Pt/AlO$_x$ sample. This is nearly one order of magnitude smaller than the recently reported values $4 \times 10^7$ A cm$^{-2}$ for the collinear antiferromagnet/Pt devices[4], and comparable to the values reported for antiferromagnet/ferromagnet[27] and collinear antiferromagnet devices[3,28]. The increase of the device temperature due to Joule heating at the maximum current pulse is estimated to be 38 K (Supplementary Fig. S10). The temperature of the device is still substantially lower than the Néel temperature ($\approx$ 430 K) even when $J_{write}$ is on, thus the Joule heating effect cannot be the predominant reason behind the observed switching. Furthermore, we performed the same zero-field current-induced switching measurements at lower temperatures (200 and 250 K), which display similar behaviors of switching with slightly larger threshold currents (Supplementary Fig. S11).

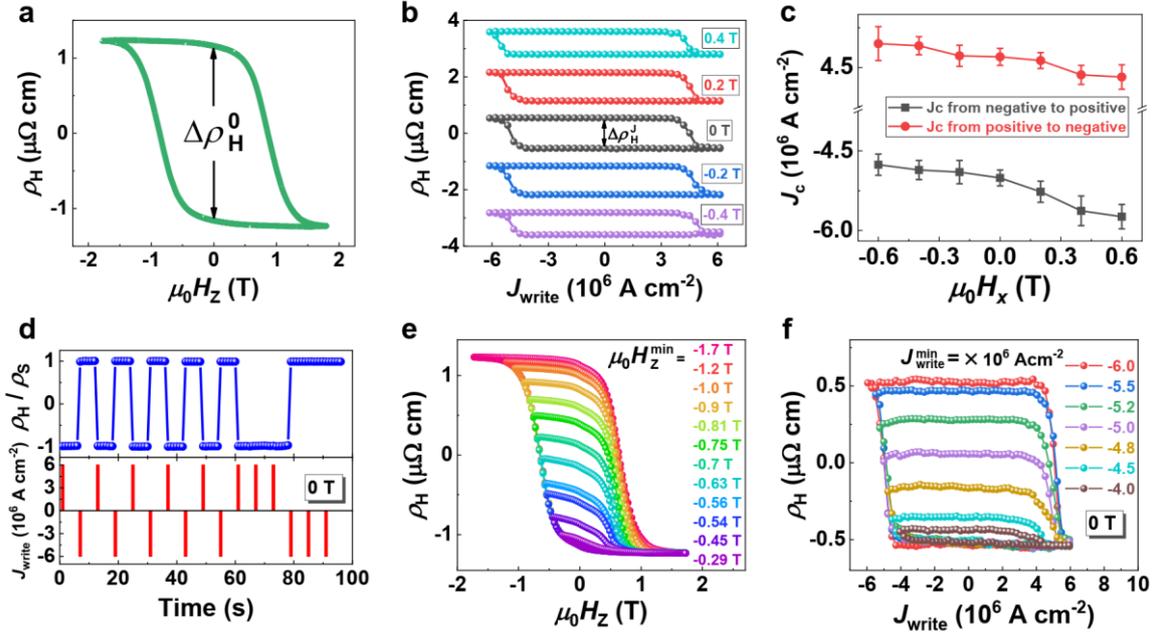

**Fig. 2 | Zero-field current-induced switching of the antiferromagnetic states in the Mn$_3$Sn device**. **a**, Anomalous Hall resistivity $\rho_H$ dependence on $\mu_0 H_Z$ for the Mn$_3$Sn device at room temperature. **b**, $\rho_H$ versus $J_{write}$ at $\mu_0 H_x = 0, \pm 0.2$ T, $\pm 0.4$ T for the Mn$_3$Sn device at room temperature. **c**, The critical current density $J_c$ as a function of $\mu_0 H_x$ for positive and negative current-sweeping directions, respectively, of the $\rho_H$ vs $J_{write}$ loops. **d**, The normalized $\rho_H$ (top) and $J_{write}$ (bottom) for the Mn$_3$Sn device at room temperature for a series of applied positive and negative $J_{write} = 6 \times 10^6$ A cm$^{-2}$, where $\rho_H$ was measured at $I_{read} = 0.1$ mA after each writing current pulse. **e**, **f**, $\rho_H$ versus $\mu_0 H_Z$ and $J_{write}$ loops for the Mn$_3$Sn device at room temperature. The minimum $\mu_0 H_Z^{min}$ and $J_{write}^{min}$ determine the magnitude of the field- and current-driven Hall resistivity switching.

Our experiments confirm that the deterministic switching of the Mn$_3$Sn devices is due to the current-induced torque exerted on the non-collinear antiferromagnetic spin texture. This reproducible bipolar switching can act as an antiferromagnetic memory. The alternating current pulses ($|J_{write}| > J_c$) along opposite directions switch the antiferromagnetic state back and forth reproducibly (Fig. 2d), indicating its reliable controllability. Remarkably, our Mn$_3$Sn device can provide multilevel signals by varying the magnitude of $\mu_0 H_Z$ or $J_{write}$. With a fixed maximum positive magnetic field (Fig. 2e) or writing current (Fig. 2f), the change of $\rho_H$ increases with increasing the magnitude of $\mu_0 H_Z^{\min}$ or $J_{\text{write}}^{\min}$ (Section III in Methods). This multi-level plasticity controlled by electrical currents at zero magnetic field suggests a great potential of Mn$_3$Sn in neuromorphic computing[21,27,29,30].

To compare the efficiency of the readout signal driven by an electrical current with that of other heavy metal/magnet bilayer structures, we defined the readout efficiency as $\xi = \Delta\rho_H^J/J_c$. In the case of our Mn$_3$Sn device, $\xi$ is approximately $2.4 \times 10^{-13}$ $\Omega$ cm$^3$/A (marked with a red star in Fig. 3), which is close to those of the ferromagnetic materials such as Co$_2$MnGa, CoFeB, *etc.*[31-38] but one to three orders magnitude larger than those of ferrimagnets, collinear antiferromagnets, and Mn$_3$Sn/Pt devices[13,39-41]. Interestingly, a scaling law of $\xi$ with the magnetization $M$ is observed for collinear antiferromagnets, ferrimagnets and ferromagnets, as indicated by the shaded region in Fig. 3. This is because the anomalous Hall resistance is proportional to the magnetization, whereas the critical switching current is insensitive to the magnetization in the case of the current-induced magnetization switching dominated by the spin Hall effect in heavy metal/magnet bilayer systems[42]. The large $\xi$ for Mn$_3$Sn is due to the strong anomalous Hall effect originated from the non-zero Berry curvature in momentum space[7,8]. The readout efficiency obtained in our pure Mn$_3$Sn is one order of magnitude higher than that in Ref. 13, probably because our Mn$_3$Sn devices have the higher inversion asymmetry and without a current being injected into the heavy metal layer. We can thus achieve a strong readout anomalous Hall signal driven by a small writing current of $5 \times 10^6$ A cm$^{-2}$ in our Mn$_3$Sn device.

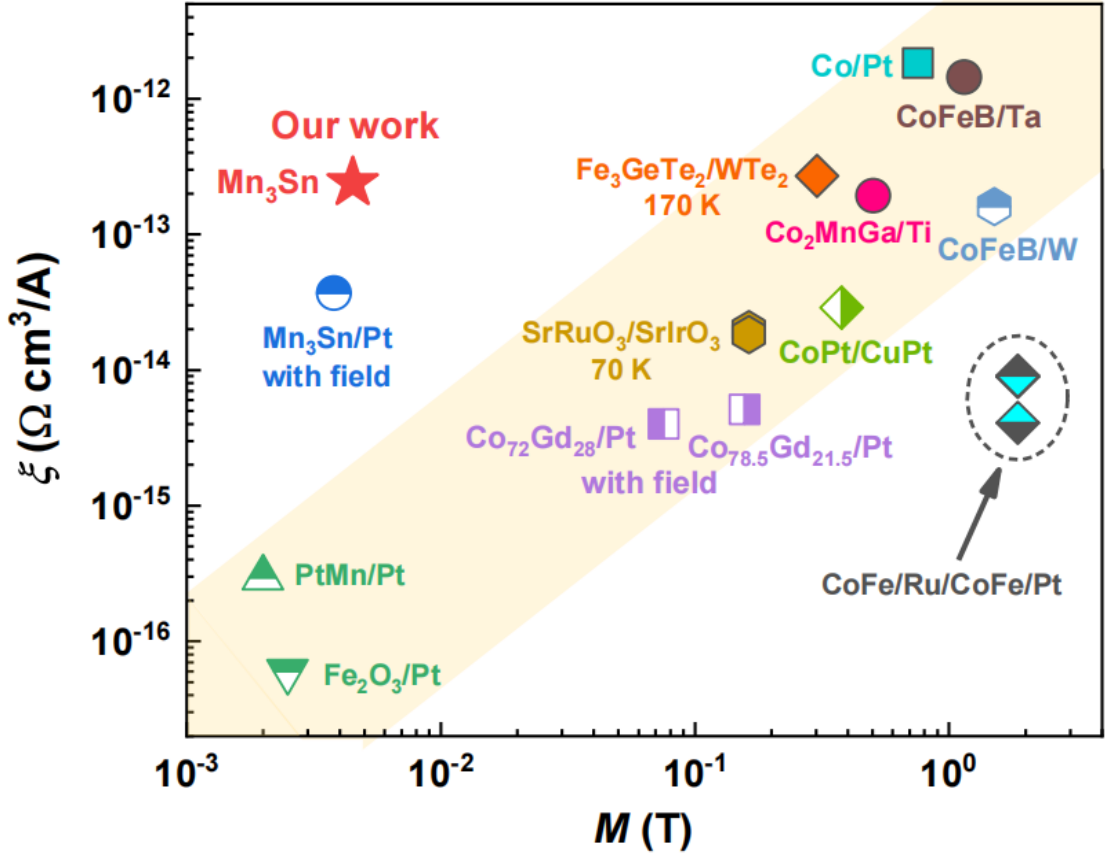

**Fig. 3 | Readout efficiency as a function of magnetization**. Double-logarithmic plot of the readout efficiency $\xi = \Delta\rho_H^J/J_c$ as a function of the magnetization $M$ for various materials, including ferromagnets (Co/Pt[31], CoFeB/Ta[32], CoFeB/W[33], CoPt/CuPt[34], CoFe/Ru/CoFe/Pt[35], Fe$_3$GeTe$_2$/WTe$_2$[36], Co$_2$MnGa/Ti[37], and SrRuO$_3$/SrIrO$_3$[38]), ferrimagnets (Co$_x$Gd$_{1-x}$/Pt[39]), collinear antiferromagnets (PtMn/Pt[40], and Fe$_2$O$_3$/Pt[41]), and Mn$_3$Sn/Pt[13], at various temperatures. The red star indicates our Mn$_3$Sn device at room temperature. The yellow shaded region highlights the scaling law of $\xi$ with $M$.

Now we illustrate the possible mechanisms of intrinsic non-collinear spin-orbit torques that induce the deterministic all-electrical switching in our Mn$_3$Sn device. We argue that the current-induced switching requires an inversion asymmetry in our polycrystal device, for the following reasons.

In our Mn$_3$Sn polycrystal, the measured non-zero anomalous Hall signal implies that the different configurations in our samples are not compensated. To understand this, we note that any crystal grain in the polycrystalline Mn$_3$Sn can be decomposed into the $z$-$x$, $x$-$y$, and $y$-$z$ configurations, according to the direction of kagome plane (Fig. 4a). The octupole rotates only in the kagome plane, so the out-of-plane magnetic field $H_z$ can only switch configurations $z$-$x$ and $y$-$z$. Considering the weak tunneling between the kagome layers, $I_x$ is not expected to be able to switch configuration $y$-$z$ effectively, i.e., the in-plane current $I_x$ can only switch configurations $z$-$x$ and $x$-$y$. The anomalous Hall signal depends on the $z$-component of the octupole, so $V_{\text{AHE}}$ can be used to read out only configurations $z$-$x$ and $y$-$z$. This infers that the Hall response of the $\mu_0 H_z$-hysteresis is around 2 to 2.3 times (considering that the octupole may relax at positions within ±30 degrees from the full polarization) to that of the $J_{\text{write}}$-hysteresis, close to the data in Figs. 2a and 2b. According to the above discussion, only configuration $z$-$x$ in Fig. 4a can be switched by the current and measurable to the anomalous Hall effect.

More importantly, the $z$-$x$ configuration of single crystal Mn$_3$Sn requires an inversion asymmetry to be deterministically switched. Because the octupole is a pseudo vector, which is invariant under inversion, so it does not reverse as the current changes sign if there is inversion symmetry (see also the symmetry and microscopic analysis in Supplementary Sections S13 and S14). The occurrence of the inversion asymmetry in polycrystalline Mn$_3$Sn was confirmed by our nonlinear Hall measurements. The inversion asymmetry can induce Rashba-like spin-orbit coupling, which can convert the injected electric current into spin currents or spin accumulations. The spin accumulations, when not aligned with the local Mn moments, can induce intrinsic non-collinear spin-orbit torques to rotate the Mn moments. The octupole is defined by the Mn moments in the Mn$_3$Sn magnetic structure with inverse triangular spin structure (Supplementary Sections S15). Thus, in this sense, it is switchable by the current, and its $z$ component leads to a measurable anomalous Hall signal.

To verify our speculations regarding the microscopic mechanism behind the observed switching, we performed numerical simulations for the octupole polarization. For a given electrical current $I_x$, the simulation starts with an initial magnetic structure of the Mn moments $\mathbf{m}_{ia}$, which behave as local Zeeman fields on the electrons described by a s-d model. Using the linear-response theory, the local spin accumulations induced by the current in the presence of the Rashba spin-orbit coupling are calculated. The effective magnetic field of the spin accumulations is then converted into magnetic torques $\mathbf{T}_{ia}$ in the Landau-Lifshitz-Gilbert equation to generate a new magnetic structure. The above steps are iterated until the magnetic structure converges to yield the octupole moment for the given injected current (see details in Methods Section IV).

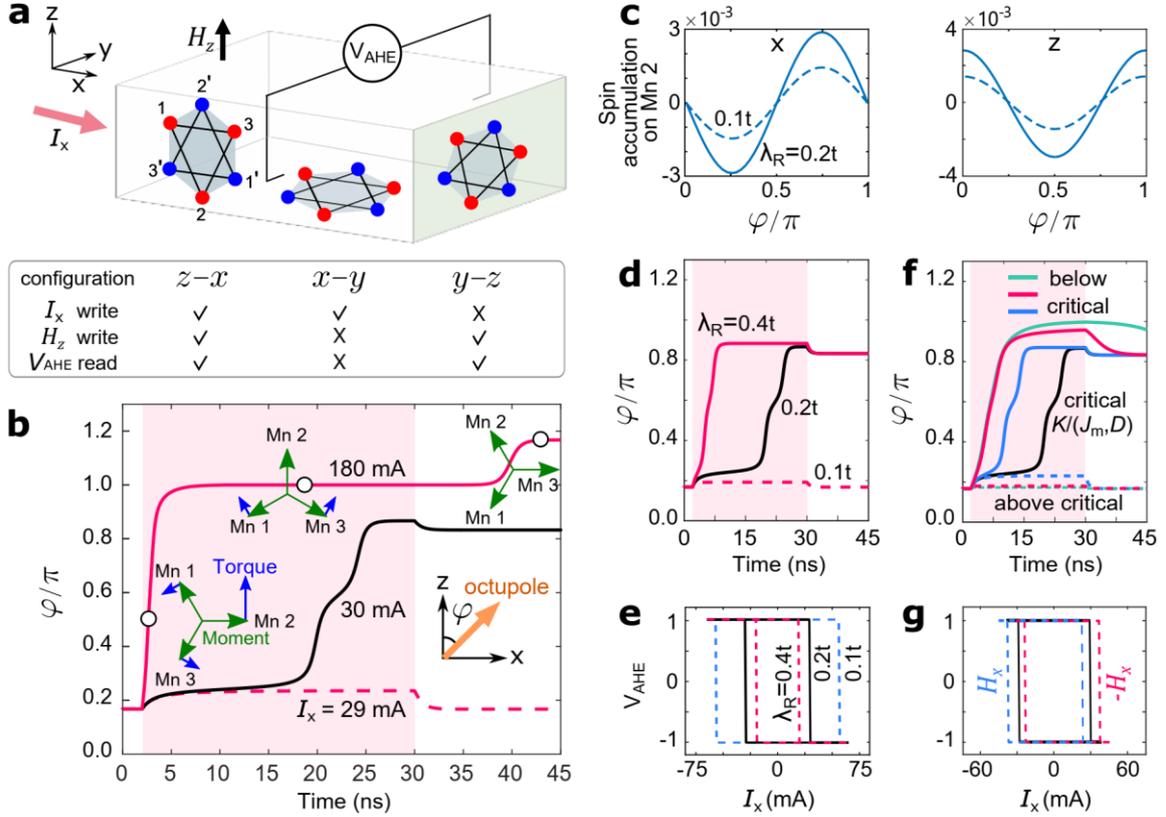

**Fig. 4 | Numerically simulated switching of the Mn$_3$Sn octupole.** The spin-orbit coupling converts the current $I_x$ (experimental $J_{\text{write}}$) into spin accumulations, which exert intrinsic non-collinear spin-orbit torques to rotate the Mn moments and octupole. **a**, The kagome lattices in polycrystal can be decomposed into three configurations $z$-$x$, $x$-$y$, and $y$-$z$. $I_x$ can switch configurations $z$-$x$ and $x$-$y$ and $V_{\text{AHE}}$ can read out configurations $z$-$x$ and $y$-$z$, so the simulation focuses on configuration $z$-$x$. **b**, Simulated octupole angle $\varphi$ versus time, driven by a $I_x$ pulse with duration of 2- 30 ns (pink shadowed region), for $I_x$ below (29 mA), at (30 mA), and well above (180 mA) the critical current. For $I_x$=180 mA, the insets show the dynamics of the torques (blue arrows) exerting on the Mn moments (green arrows), at three stages (hollow circles), as $\varphi$ starts from $\pi/6$ (an easy axis), then is switched to $\pi$ (normal to $I_x$) after turning on $I_x$, and finally relaxes to $7\pi/6$ (another easy axis) after turning off $I_x$. **c**, Simulated $x$- and $z$-direction spin accumulations on Mn 2 (denoted in panel **a**), as a function of $\varphi$, for different Rashba spin-orbit coupling $\lambda_R$ (Eq. 2 in Methods). **d**, The same as the 30 mA case in panel **b**,

but also for $\lambda_R$ below and above the critical value. **e**, Simulated hysteresis loops of the Hall response $V_{AHE}$ driven by $I_x$, for different $\lambda_R$. **f**, The same as the 30 mA case in panel **b**. Combinations of the magnetic-structure parameters $K/(J_m, D)$ (Eq. 1 in Methods) below (solid curves) and above (dashed curves) the critical combination are also considered. **g**, The same as the $\lambda_R = 0.2t$ case in panel **e**, The presence of the $x$-axis magnetic field $H_x$ is also considered. More simulation results by changing $\lambda_R$, $K/(J_m, D)$, and initial states can be found in Supplementary Section S16.

Figure 4b shows the simulated octuple angle $\varphi$ as a function of time at different current intensities, which demonstrates that the current must be sufficiently strong to drive a switching. Its insets also show the microscopic dynamics in terms of the calculated torques exerting on the Mn moments at different stages of a successful switching for $I_x$ well above the critical value. The noncollinear antiferromagnetic structure tends to maintain during the switching and the rotation of the Mn moments (collectively as octupole) is determined by the sum of the nutation tendencies of the sublattice moments driven by the torques. Figures 4c-e show that a larger Rashba spin-orbit coupling $\lambda_R$ leads to an enhanced spin accumulation, faster switching, and smaller critical current. Figure 4f shows that increasing/decreasing the magnetic-structure parameters $K/(J_m, D)$ leads to a failed/faster switching, where $J_m$ and $D$ stabilize the inverse 120° triangular structure, whereas $K$ introduces a small deviation to exert torques that relax the octupole to one of the six stable positions ($\varphi = \pi/6 +$ integer times of $\pi/3$). In other words, the switching requires the torques from the current-induced spin accumulations to overcome the stable-position-favored torques due to the magnetic structure, which favours the six stable positions. Figure 4g shows that the hysteresis loop can be shifted by the $x$-direction magnetic field of $\pm 0.0012$ T (the theoretical values are usually orders of smaller than those in the experiments[9,13], probably because the external magnetic field is screened in the realistic polycrystals). More importantly, the simulations confirm that, without external magnetic fields and heavy metals, the all-electrical switching of the non-collinear antiferromagnet can be achieved, owing to the intrinsic non-collinear spin-orbit torques from the current-induced local spin accumulations in Mn₃Sn itself.

Our findings promise potential applications of Mn₃Sn in information technologies. Specifically, the all-electrical control of the binary and multilevel states with large Mn₃Sn readout signals could act as the building block for magnetic random access memory and artificial synapses, respectively. Furthermore, the in-memory computing will be also achieved utilizing the all-electrical control of topological non-collinear antiferromagnets.

**Acknowledgments**

This work was supported by the National Key R&D Program of China (grant no. 2017YFA0303400), the National Natural Science foundation of China (grant No.s 11474272, and 11925402), the Chinese Academy of Sciences (grant No.s QYZDY-SSW-JSC020, XDB28000000, and XDB44000000), the Beijing Natural Science Foundation (grant No. 2212048), the Beijing Natural Science Foundation Key Program (grant No. Z190007), Guangdong (2016ZT06D348, 2020KCXTD001), Shenzhen (G02206304, G02206404, ZDSYS20170303165926217, JCYJ20170412152620376), and Center for Computational Science and Engineering of SUSTech. The authors thank Prof. Hong Ding for fruitful discussions.

**Author contributions**

K.W. conceived the work. Y.D. and X.L. grew the films, fabricated the devices and carried out the electrical transport measurements. N.J and C.S. measured the magnetic properties of the samples. Y.D., X.L. and K.W. analyzed the data. Y.C., Z.D. and H.L. performed theoretical calculations. Y.D., X.L., H.L. and K.W. wrote the manuscript. All the authors discussed the results and commented on the manuscript.

**Competing interests**

The authors declare no competing interests.

**Data and materials availability**

Data for all graphs presented in this study are available upon request from the corresponding authors.

# Methods

**I. Sample and device fabrication.** The samples used for the current-induced switching measurements, which consist of Mn$_3$Sn (50)/AlO$_x$ (2) (the thicknesses shown in parentheses are in nanometers), were grown on thermally oxidized Si substrates, the top AlO$_x$ layer was used as the capping layer. For comparison, reference samples consisting of Ru (3)/Mn$_3$Sn (50)/AlO$_x$ (2) and Ru (3)/Mn$_3$Sn (50)/Pt (8)/AlO$_x$ (2) were also deposited on Si/SiO$_2$. The Mn$_3$Sn, Ru, and Pt layers were deposited at room temperature using a DC magnetron sputtering system with a base pressure of less than $5 \times 10^{-9}$ Torr (at rates of approximately 0.05, 0.01 and 0.02 nm/s, power of 30 W and Ar gas pressure of 0.8 mTorr). Next, the AlO$_x$ layers were grown using an RF magnetron sputtering system with a power of 80 W and Ar gas pressure of 2 mTorr. The Mn$_3$Sn (50)/AlO$_x$ (2), Ru (3)/Mn$_3$Sn (50)/AlO$_x$ (2) and Ru (3)/Mn$_3$Sn (50)/Pt (8)/AlO$_x$ (2) samples were then annealed at 450 °C for 1 hour using a vacuum annealing furnace (F800 − 35, East Changing Technologies, China) at a base pressure of $5 \times 10^{-7}$ Torr. The samples were patterned into Hall bar devices with a current channel width of 10 μm using the standard photolithography and Ar-ion etching.

**II. Microstructural characterization.** The structural analysis was carried out via X-ray diffraction using a monochromated Cu K$_\alpha$ radiation ($\lambda = 1.54056$ Å). The sample for cross-sectional scanning transmission electron microscopy characterization was prepared using a Zeiss Auriga focused ion beam system. The high-angle annular dark-field (HAADF) Z-contrast images, bright-field images, and EDX mappings were acquired using a spherical aberration-corrected FEI Titan Cubed Themis 60-300 operated at 200 kV.

**III. Magnetic and magnetotransport measurements.** The magnetic measurements of our polycrystalline Mn$_3$Sn samples were carried out through MCD spectroscopy using a wavelength tunable laser via a supercontinuum white light source equipped with a monochromator. The reflected light from the sample was detected using a Si photodetector. The reflectance and MCD signals were obtained using two lock-in amplifiers with reference frequencies of 177 Hz and 50 kHz, respectively. Moreover, a Quantum Design vibrating sample magnetometer (VSM), Physical Property Measurement System (PPMS) with a closed-cycle helium cryostat was used to investigate the magnetic properties of our Mn$_3$Sn thin film. A reading current of $I_{\text{read}} = 0.1$ mA was used to measure the Hall resistivity $\rho_\text{H}$ of the Mn$_3$Sn Hall bar devices, as shown in Fig. 1d. In the current-induced switching measurements, we employed a writing current pulse (with current density $J_{\text{write}}$) with a duration of 50 ms and subsequently, we applied $I_{\text{read}}$ to read out the Hall voltage after waiting for 500 ms.

The angle dependence of magnetoconductivity and planar Hall effect were obtained using the horizontal rotator option of a commercial magnetoelectrical measurement system. The electrical measurements were conducted using a cryogenic probe station (Lake Shore Cryotronics, Inc). For the $\rho_H - \mu_0 H_Z$ loop measurements shown in Fig. 2e, we first initialized the antiferromagnetic states by applying a saturation field of 1.8 T, and we then scanned $\mu_0 H_Z$ from 1.8 T to $\mu_0 H_Z^{\min}$ and back to 1.8 T. Similarly, for the $\rho_H - J_{\text{write}}$ loop measurements shown in Fig. 2f, we first initialized the antiferromagnetic states using a positive current pulse (with a current density of $6 \times 10^6$ A cm$^{-2}$ larger than the switching current density $J_c$). We then scanned $J_{\text{write}}$ from $6 \times 10^6$ A cm$^{-2}$ to $J_{\text{write}}^{\min}$ and back to $6 \times 10^6$ A cm$^{-2}$ at zero magnetic field.

For the nonlinear Hall measurements, a sinusoidal current with a frequency of 133.33 Hz along the longitudinal direction was applied to the Hall devices using standard lock-in techniques (Stanford Research Systems Model SR830), the second harmonic voltages were detected in a direction perpendicular to the current. All nonlinear Hall measurements were carried out at room temperature without external magnetic field. The relationship between the nonlinear Hall voltage and current was obtained by gradually sweeping the amplitude of the sinusoidal current.

**IV. Simulating current-induced switching of the non-collinear antiferromagnet.**

To understand the microscopic mechanism of the all-electrical switching of the non-collinear antiferromagnet, we performed a numerical simulation of the octupole polarization direction $\boldsymbol{O} = (O_x, O_z)$, which can be determined from the direction of the magnetic moment $\mathbf{m}_{i2}$ (assumed to be the same for different unit cells $i$ in the simulation) of the second Mn atom in the unit cell as

$$O_x = (\mathbf{m}_{i2})_x, O_z = -(\mathbf{m}_{i2})_z,$$

where $i$ indexes the unit cells. This expression is simplified from the full definition in Supplementary Section S15. The simulation procedures are as follows.

**Step 1** Assume an initial magnetic structure of the Mn magnetic moments $\mathbf{m}_{ia}$ (Eq. 1 in Methods), $a = \{1, 2, 3\}$ denotes the three Mn atoms in each unit cell.

**Step 2** Use the real-space form of an s-d model (Eq. 2 in Methods) to describe the itinerant electrons in the local Zeeman fields exerted by the Mn magnetic moments $\mathbf{m}_{ia}$, where s for the itinerant electrons and d for the Mn magnetic moments.

**Step 3** For an input electric current $I_x$, use the linear-response theory for the current-spin correlation (Eqs. 3 and 4 in Methods) to calculate the local spin accumulations of the itinerant electrons induced by the current $I_x$.

**Step 4** Use the local spin accumulations $\mathbf{S}_a$ of the itinerant electrons to calculate the magnetic torques $\mathbf{T}_{ia}$ exerted on the Mn magnetic moments $\mathbf{m}_{ia}$ (Eq. 5 in Methods).

**Step 5** Use the magnetic torques $\mathbf{T}_{ia}$ and the Landau-Lifshitz-Gilbert equation (Eq. 6 in Methods) to calculate a new magnetic structure of $\mathbf{m}_{ia}$.

**Step 6** Return to Step 1 with the new magnetic structure.

Repeat steps 1-6 until the magnetic structure converges to yield the octupole moment in the presence of the injected current $I_x$. Any crystal grain configuration can be constructed using the three vector grain configurations of our polycrystalline Mn$_3$Sn Hall-bar devices (Fig. 4a). Here we focus on the *z-x* configuration (left-hand side in Fig. 4a) because we found that the switching accompanied by the reversing of the anomalous Hall resistivity due to the electric current occurs only in this configuration.

**Magnetic structure of Mn moments.** To understand the microscopic mechanism of the current-induced switching of the non-collinear antiferromagnetic states, we study the dynamics of the sublattice moments $\mathbf{m}_{ia}$ using the following Hamiltonian defined on a single layer of the Mn$_3$Sn kagome lattice[13] (on the $x - z$ plane in Fig. 4a)

$$H_m = D \sum_{\langle i,j \rangle} \mathbf{c} \cdot (\mathbf{m}_{i1} \times \mathbf{m}_{j2} + \mathbf{m}_{i2} \times \mathbf{m}_{j3} + \mathbf{m}_{i3} \times \mathbf{m}_{j1}) + J_m \sum_{\langle ia,jb \rangle} \mathbf{m}_{ia} \cdot \mathbf{m}_{jb} - K \sum_{ia} (\mathbf{k}_a \cdot \mathbf{m}_{ia})^2, \quad (1)$$

where $i$ and $j$ index the unit cells, $a, b = \{1, 2, 3\}$ denote the sublattices, and $\mathbf{c} = (0, 1, 0)$ is the unit vector along the y axis (Fig. 4a). This model contains only three Mn atoms in a unit cell, and we neglect inter-layer couplings because they do not provide qualitatively new effects. The Dzyaloshinskii-Moriya interaction $D$, nearest neighbour exchange interaction $J_m$, and in-plane magnetic anisotropy $K$ are assumed to be positive, and they stabilize the inverse triangular spin texture of Mn$_3$Sn. $\mathbf{m}_{ia}$ represents a unit magnetic moment on site $ia$. $\mathbf{k}_a = (\sin \phi_a, 0, \cos \phi_a)$, with $(\phi_1, \phi_2, \phi_3) = \frac{(\pi, 9\pi, 5\pi)}{6}$. The $K$ term lifts the in-plane U(1) degeneracy and fixes the sixfold symmetry, introducing six stable positions for the octupole. For the magnetic structure, we used a single unit cell with periodic boundary conditions, the same as that in the previous work[13].

**s-d model**. To establish the relationship between the injected current and the local Mn magnetic moments and calculate the current-induced local spin accumulations, we employed an s-d model from the previous work[40]

$$H_{sd} = t \sum_{\langle ia,jb \rangle n} c^\dagger_{ian} c_{jbn} + J_{sd} \sum_{iann'} \mathbf{m}_{ia} \cdot \boldsymbol{\sigma}_{nn'} c^\dagger_{ian} c_{ian'} + \Lambda \sum_{in, a=\{1,2,3\}} c^\dagger_{ian} c_{ian}$$

$$+ \lambda_R \sum_{\langle ia,jb \rangle nn'} i \left( \hat{\mathbf{y}} \times \hat{\mathbf{d}}_{ia,jb} \right) \cdot \boldsymbol{\sigma}_{nn'} c^\dagger_{ian} c_{jbn'}, \quad (2)$$

where $c_{ian}$ is the annihilation operator of the electron highly localized on site $ia$ (notice that sublattice index $a = \{1,2,3,1',2',3'\}$ here corresponds to a monolayer AB stack $Mn_3Sn$ structure and we only need to calculate the local spin accumulations on sublattice 1,2,3 for the simulation) with spin $n = \{\uparrow, \downarrow\}$, $t$ is the kinetic energy, $\Lambda$ is the energy difference between the $\{1,2,3\}$ and $\{1',2',3'\}$ layers, $J_{sd}$ is the exchange interaction between the local moments and the spins of itinerant electrons, $\boldsymbol{\sigma}$ is the spin operator of the electrons, $\hat{\mathbf{d}}_{ia,jb}$ is the unit vector from site $ia$ to $jb$, and $\lambda_R$ measures the Rashba-type spin-orbit coupling that breaks the mirror time reversal symmetry and inversion symmetry. By performing a Fourier transformation, the Hamiltonian in Eq. 2 can be converted from the real space to the $k$ space. For each $k$ point in the $k$ space, the Hamiltonian is a $k$ labeled $12 \times 12$ matrix (See Supplementary Section S17). Diagonalizing the matrix provides us with the energy spectrum and eigenstates of the s-d model. In Supplementary Section S14, we show that a reasonable energy difference ( $\Lambda = 0.1t$, comparable with the Rashba spin-orbit coupling) does not cause qualitative change to our switching mechanism. To simplify the illustration of the switching mechanism, we assume $\Lambda = 0$ in the simulations shown in our main text.

**Current-induced spin accumulations and torques**. Using the linear-response theory, the Kubo formula for the local spin accumulations on sublattice $a$ in response to the electric field (i.e., the injected write current) along the $\beta = \{x, y, z\}$ direction can be written as

$$\delta \langle \mathbf{s}_a \rangle_d = -\frac{e\hbar}{2} \tau E^\beta \sum_n \int \frac{d\mathbf{k}}{(2\pi)^2} \frac{\partial f}{\partial \varepsilon_n} v^\beta_{nn} \sigma^a_{nn}, \quad (3)$$

$$\delta \langle \mathbf{s}_a \rangle_{od} = -\frac{e\hbar^2}{2} E^\beta \sum_{m \neq n} \int \frac{d\mathbf{k}}{(2\pi)^2} (f_m - f_n) \times \mathrm{Im} \left[ \frac{v^\beta_{mn} \sigma^a_{nm}}{(\varepsilon_m - \varepsilon_n)^2} \right], \quad (4)$$

where "d" stands for diagonal, and "od" stands for off-diagonal. The electric field can be derived from the injected current $I$ as $E = \frac{m^* I}{\omega d n_{3D} e^2 \tau}$, where $m^*$ is the electron effective mass, $\omega$ and $d$ are the width and thickness of the device, respectively, $n_{3D}$ is the carrier density, $e$ is the elementary charge, $\tau$ is the relaxation time, $f$ is the Fermi-Dirac distribution function, $\varepsilon$ is the energy of state, $v$ denotes the velocity operator, and $\boldsymbol{\sigma}^a = (\sigma_x^a, \sigma_y^a, \sigma_z^a)$ represents the local spin operator on sublattice $a$. In Fig. 4c, the unit of the spin accumulation is $s_0 = I_x\, m^*/2\, n_{3D}\, e\, a_L$, with $a_L$ being the kagome plane lattice constant. In the calculation of the local spin accumulations, we used $50 \times 50$ unit cells, which is sufficient to yield convergent results. The local spin accumulations on the sublattice produce a field-like torque[2]

$$\mathbf{T}_{ia} = -|\gamma|\mathbf{m}_{ia} \times \mathbf{H}_a^I, \text{ with } \mathbf{H}_a^I = -\frac{J_{sd}}{M_S} \frac{\mathbf{S}_a}{\frac{\hbar}{2}}, \quad (5)$$

where $\mathbf{S}_a = A_U \times \delta \langle \mathbf{s}_a \rangle_d$ is the induced local spin accumulations on the sublattice $a$ of a unit cell with area $A_U$, and $M_S = 3\mu_B$ is the saturation magnetic moment of an Mn atom. $\delta \langle \mathbf{s}_a \rangle_{od}$ are not included because in the presence of the Rashba spin-orbit coupling, the spin accumulation calculated from Eq. 4 on an Mn atom is always opposite to that of its diagonal counterpart in the other kagome layer (see the symmetry analysis in Supplementary Section S14). In other words, the off-diagonal spin accumulations do not contribute to the switching.

**Landau-Lifshitz-Gilbert (LLG) equation**. The LLG equation describing the dynamics of the magnetic moments on the sublattices of a single layer of the kagome lattice in Mn$_3$Sn is

$$\dot{\mathbf{m}}_{ia} = -|\gamma|\mathbf{m}_{ia} \times \mathbf{H}_{\text{eff},ia} + \alpha \mathbf{m}_{ia} \times \dot{\mathbf{m}}_{ia} + \mathbf{T}_{ia}, \quad (6)$$

where the suffix $i$ represents a unit cell, and $a = \{1, 2, 3\}$ indexes a sublattice. The effective magnetic field is given by $\mathbf{H}_{\text{eff},ia} = -M_S^{-1} \delta H_m / \delta \mathbf{m}_{ia}$, with $H_m$ being the Hamiltonian of the magnetic structure (Eq. 1). The first and second terms in the right-hand side of Eq. 6 represent the gyroscopic torque and Gilbert damping torque, respectively, $\gamma$ ($\gamma < 0$) is the electron gyromagnetic ratio, and $\alpha$ is the Gilbert damping coefficient. The third term $\mathbf{T}_{ia}$ is the spin torque (Eq. 5) acting on the moments of the corresponding sublattices. Using this equation, we can simulate the dynamics of the magnetic moments on the sublattices to understand how $J_{\text{write}}$ influences the polarization of the octupole and determines the sign of the anomalous Hall resistivity, as shown in Fig. 4b. In Fig. S16 we show that our switching mechanism is not influenced by initial state of magnetic structure, i.e., the final state of the switching is determined by the direction of injected current. To numerically solve the LLG equation, we used ODE23 Matlab function, in which the time scale is self-adapted, i.e., it changes depending on the precision requirement.

**Simulation parameters.** In Fig.4, unless specified on the curves, the simulation parameters are $t = 0.25$ eV, $J_{sd} = 0.375$ eV, $\Lambda = 0$, $\lambda_R = 0.2$ t, $\tau = \hbar/2\Gamma$ with $\Gamma = 1.25$ meV, $m^* = 4.05 \times 10^{-31}$ kg, $\omega = 15$ μm, $d = 40$ nm, $n_{3D} = 6 \times 10^{23}$ cm$^{-3}$, $T = 290$ K, $J_m = 23$ meV, $D = 1.6$ meV, $K = 0.17$ meV, $\alpha = 0.003$[13,43]. Using these parameters in the simulations, a critical switching current of $I = \pm 30$ mA is obtained, which corresponds to the current density $J_{\text{write}} \approx 5 \times 10^6$ A cm$^{-2}$ in the measurement. The $K/(J_m, D)$ values used in Fig. 4f are 0.085/(23,1.6), 0.17/(46,1.6), and 0.17/(23,3.2) for the above-critical curves and 0.34/(23, 1.6), 0.17/(11.5, 1.6), and 0.17/(23,0.8) for the below-critical curves, all in units of meV.